# Power Networks SCADA Communication Cybersecurity, A Qiskit Implementation

Hillol Biswas
Department of Electrical and Computer Engineering
Democritus University of Thrace,
XANTHI, Greece

*Abstract*— The cyber-physical system of electricity power networks utilizes supervisory control and data acquisition systems (SCADA), which are inherently vulnerable to cyber threats if usually connected with the internet technology (IT). Power system operations are conducted through communication systems that are mapped to standards, protocols, ports, and addresses. Real-time situational awareness is a standard term with implications and applications in both power systems and cybersecurity. In the plausible quantum world (Q-world), conventional approaches will likely face new challenges. The unique art of transmitting a quantum state from one place, Alice, to another, Bob, is known as quantum communication. Quantum communication for SCADA communication in a plausible quantum era thus obviously entails wired communication through optical fiber networks complying with the typical cybersecurity criteria of confidentiality, integrity, and availability for classical internet technology unless a quantum internet (qinternet) transpires practically. When combined with the reverse order of AIC for operational technology, the cybersecurity criteria for power networks' critical infrastructure drill down to more specific sub-areas. Unlike other communication modes, such as information technology (IT) in broadband internet connections, SCADA for power networks, one of the critical infrastructures, is intricately intertwined with operations technology (OT), which significantly increases complexity. Though it is desirable to have a barrier called a demilitarized zone (DMZ), some overlap is inevitable. This paper highlights the opportunities and challenges in securing SCADA communication in the plausible quantum computing and communication regime, along with a corresponding integrated Qiskit implementation for possible future framework development.

## I. Introduction

As the smart grid entails a bi-directional flow of power and data, cybersecurity is also paramount for protecting the grid, depending upon the size and complexity. Globally, almost all electrical power transmission substations adhere to domain-specific standards. SCADA, since its inception [1], historically has evolved from monolithic to distributed to networks with applicable connection to RTU/load despatch center [2]. The IEC 61850 standard encompasses communication networks and systems in substations, covering various components for specific application areas. The substation automation system, based on IEC 61850, encompasses communication, signalling, and functionality. The exchanges guarantee that the correct procedures are followed and flow from and to a specific source and destination address. For substation control, security, and monitoring, the IP address tracks the appropriate ports where packets are transmitted at predetermined times, formats, and values. When installed in either real-world scenarios or lab-based experimental setups, the records are captured by the network protocol analyzer Wireshark, which provides valuable insights into the complexities of the IEC 61850-based communication mechanism. However, the ability to stream data quickly and in real time inevitably poses a corresponding cyber risk. By transmitting and communicating with the remote terminal unit (RTU) or load dispatch center, other standards, such as IEC 60870–5–101 and 104, are also adhered to in parallel. Moreover, the cybersecurity paradigm adheres to, among other things, the IEC 62351 series for Power systems management and associated information exchange—data and communications security.

Therefore, the complexity is multifaceted depending upon the size, and with the growing cybersecurity landscape specifically for this critical infrastructure domain, emerging threats and vulnerabilities are paramount. While cyberattacks like denial of service (DOS)/ distributed denial of service (DDOS), and false data injection (FDI), are no new in this domain, there have been instances of emerging malware targeting power networks with growing sophistication, allegedly with the ability to exploit specific protocols, such as IEC 101 or 104, with the intent of controlling substation equipment, viz. circuit breakers [3]. The cybersecurity landscape has witnessed numerous historical attacks worldwide targeting electricity networks. It is no wonder that, with the growing sophistication in technology, corresponding threats and vulnerabilities will likely emerge in response. With the advent of quantum computing, it also branches out into associated fields, such as quantum communication, quantum machine learning, and many others. Numerous literary works are emerging in the current research setting of quantum endeavors, raising a pertinent question about the communication system and corresponding cybersecurity regime of the critical infrastructure of power networks, a crucial building block of modern civilization. MITRE has both ATT and CK [4] Provide updated tactics, techniques, and procedures (TTP) guidance, including industrial control systems (ICS). However, SCADA power networks' communication and cybersecurity paradigms entail customized and specific ecosystems. As threats emerge with the advent of machine learning, the MITRE ATLAS [5] navigator cautions about specific novel vulnerabilities that are transpiring simultaneously. The current research in the quantum computing regime is potentially encouraging, so the apparent question arises: how would the corresponding cybersecurity aspects in this novel endeavor be addressed?

## II. Related Works

It has long been speculated that if a quantum computer were to emerge, current communication protocols and standards would likely be insufficient to provide an effective defense against both conventional and quantum attacks. NIST's work on post-quantum cryptography is one proactive direction for incorporating research in this area. However, quantum communication encompasses numerous additional directions, including quantum cryptography, teleportation, entanglement, quantum key distribution, and quantum hacking, each with its corresponding cybersecurity implications.

### A. Quantum Communication and Cryptography

Distribution of quantum keys QKD offers a way for authorized users named "Alice," the sender, and "Bob," the recipient, to exchange a secret key [5]. However, Eve, the eavesdropper's appearance, is also included in the scenario.

Bennet et al. first proposed, based on an EPR experiment, that a particle's quantum state may be transferred onto another particle—a process known as quantum teleportation—as long as no information about the state is obtained during the transition [6]. Further, in the pioneering work, Bownmeester et al. (1996) [7] demonstrated experimentally that the two most difficult tasks for any experimental realization are creating and measuring entangled states, which are required for teleportation. The experiment involved transferring the polarization state of one photon onto another using pairs of polarization-entangled photons created via pulsed down-conversion and two-photon interferometric techniques. However, this method is not the only one that allows teleportation. Zhang et al. [8] Provided an experimental demonstration of the original proposal by Bennett et al., a single-qubit system, and further demonstrated the experimental execution of quantum teleportation in a two-qubit composite system. In the experiment, they created and used a six-photon interferometer to teleport two photons in any arbitrary polarisation state. The observed teleportation fidelity values for various initial states are significantly higher than the state estimation limit of 0.40 for the two-qubit system.

Quantum teleportation, which uses the physical resource of entanglement, is a fundamental primitive in many quantum information tasks and a crucial component of quantum technologies. It is essential to the ongoing development of quantum networks, quantum computing, and quantum communication. Pirandola et. al., reviews the fundamental theoretical concepts underlying quantum teleportation and its variation protocols [9] as

i) Quantum repeaters and entanglement swapping

ii) Networks of Quantum Teleportation

iii) Quantum computing and quantum gate teleportation

iv) Port-based teleportation

Corresponding to these, the experimental set-up was ideally successful,

(1) The input state (within an appropriate alphabet) is arbitrary.

(2) A third party, let us assume Victor, provides Alice with the input state and independently confirms Bob's output state (for example, using fidelity estimation or quantum state tomography).

(3) Alice can differentiate a whole foundation of entangled states by performing a complete Bell detection.

(4) Teleportation fidelity is higher than what can be achieved with traditional measure-prepare techniques.

To transfer the quantum states of an actual particle, such as a single photon, one must consider both multilayer and two-level states, including polarization. Teleporting high-dimensional quantum states remains difficult for two reasons. The first is the creation of high-dimensional, high-quality entanglement, which is possible for quantum teleportation. The other is a deterministic high-dimensional Bell state measurement (HDBSM). Here, Hu et al. performed the HDBSM using an additional entangled photon pair and encoded the three-dimensional states using the spatial mode (path), which has been shown to have very high fidelity [10]. Summarizing, they demonstrate the capacity to simultaneously transfer and control a single object's high-dimensional state by reporting the quantum teleportation of that particle's high-dimensional qubits.

At first, manipulating qubits was difficult, but it is now routine work, either by adding more qubits or introducing d-dimensional space, thereby expanding the quantum Hilbert space. It is called a qutrit for three-dimensional space and a qudit in d-dimensional space, where d is greater than two. In order to transfer more than one bit or photon from Alice to Bob, the available dimensions must be increased. Various photonic degrees of freedom or a combination of them can be utilized to enlarge the Hilbert space.

Communication channels can be either free-space, such as satellite communication, or optical fiber-based, although other channels, like aquatic, are also candidates in this pursuit. Optical fiber links are the most appealing channel for transmitting high-dimensional quantum states in a free-space link, as the infrastructure is already established and optical fiber communication is widely utilized in daily life, such as in the Internet backbone. High-dimensional quantum states can be propagated using a variety of fiber types, including single-mode fibers (SMFs), multimode fibers (MMFs), which include few-mode fibers and higher-order mode fibers, and multicore fibers (MCFs), which are special fibers with multiple cores within the same cladding. Various fibre kinds offer an ideal answer, depending on the application. For instance, the footprint of the fibers is crucial in data centers, where there is a shortage of space and a notably large number of connections are needed. For this reason, using MCFs is an exceptionally alluring option [11].

### B. Optical Fiber Networks and SCADA Systems in Power Networks

Capacity and security are two of the most challenging issues that communication engineers must address, and quantum technologies can help. We may more effectively investigate the whole information-carrying capability of optical fibers by drastically reducing the number of photons required to encode each bit of information. Moreover, it can leverage quantum principles to enhance communication systems by encoding information in a single or small number of photons. Operators have been raising the bit rate per channel and the number of optical channels per fibre to handle this growing volume of traffic. In this section, we go over the capacity limitations of traditional fiber optic communication networks and examine how it might overcome them by encoding information in a single or small number of photons [12].

It is possible to construct memory effects to enhance an optical fiber's communication capabilities. This is accomplished by using trigger signals that have been initialized in an appropriate state to take advantage of the noise attenuation process. As long as preshared entanglement is consumed, this protocol enables the reliable transmission of (a) qubits at a fixed positive rate and (b) bits and qubits at a rate of the same order of the maximum possible in the ideal condition of no noise over arbitrarily long optical fibres [13].

Controlling light in all its degrees of freedom has recently gained popularity, as it has produced long-predicted novel light states, improved application functionality, and a contemporary toolkit for examining fundamental science. The spatial modes of light can be used to encode a large alphabet by structuring light as single photons and entangled states. This enables access to high-dimensional Hilbert spaces for basic quantum mechanical tests and enhanced quantum information processing tasks. The fundamental idea behind using spatial modes to realise high-dimensional quantum states. Starting the experiment, preparing and measuring quantum states, and utilizing these as resources for quantum information processing and imaging have all been covered, with real-world applications [14].

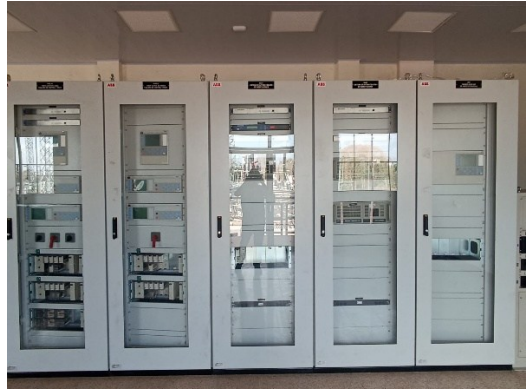

Fig. 1.   A typical substation automation system cum communication panel

Fig. 1 depicts a SCADA communication panel for the hoisting substation automation system (SAS) in a typical newly built substation. There are typically many panels, with one dedicated exclusively to communication purposes, often referred to as a SAS panel. It connects through optical fibers to various other panels of the substation, including the busbar panel, transformer panel, bus-coupler panel, and others, depending on the substation layout and communication architecture.

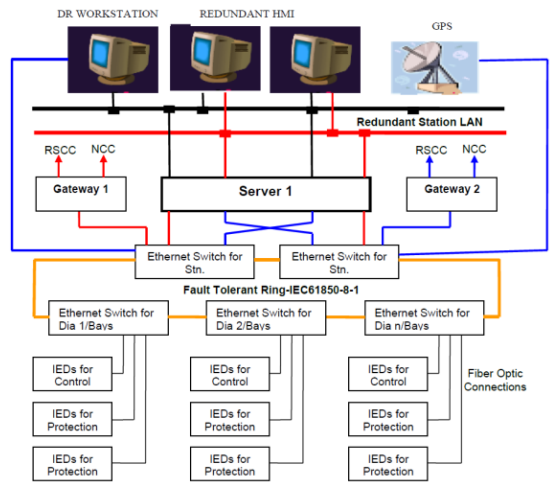

Fig. 2. A typical SCADA architecture for electrical substation communication

Fig. 2 illustrates a SCADA architecture typically used in a step-down substation, featuring components such as a human-machine interface (HMI) connected through Gateways and redundant station local area networks (LANs) via Ethernet connections, all in a fault-tolerant ring configuration that complies with IEC 61850-8-1.

A typical system architecture is defined for a substation using generic criteria. Decentralized architecture and the idea of bay-oriented, distributed intelligence will serve as the foundation for SAS. Decentralized, object-oriented functions that are situated as near to the process as feasible are required. The station's primary process information will be kept in databases that are dispersed. The basic SAS architecture comprises two levels: a station level and a bay level. All bay-level operations about control, monitoring, and protection, as well as inputs for status indication and outputs for commands, must be provided by the IEDs. There should be no need for further interposition or transducers; the IEDs should be linked straight to the apparatus. Every bay control IED must operate independently of the others, and any malfunction in the station's other bay control units cannot interfere with its operation. The communication infrastructure is being used to transfer data between the electronic devices at the bay and station levels. Fiber-optic cables are used to accomplish this, ensuring uninterrupted connection. A real-world electrical substation pcap, Fig. 3, depicts different protocols in Source and destination address with the corresponding ports and lengths in communication.

| No. | Time | Source | SPort | Destination | DPort | Protocol | Length |
|---|---|---|---|---|---|---|---|
| 46 | 06:00:07.244437 | 10.22.91.250 | 63667 | 10.22.91.22 | 502 | TCP | 54 |
| 47 | 06:00:07.351723 | Ge_0c:25:14 | | IecTc57_01:00… | | GOOSE | 214 |
| 48 | 06:00:07.385210 | 10.22.91.250 | 63667 | 10.22.91.22 | 502 | Modbus/T… | 66 |
| 49 | 06:00:07.386273 | 10.22.91.22 | 502 | 10.22.91.250 | 63667 | TCP | 60 |
| 50 | 06:00:07.386659 | 10.22.91.22 | 502 | 10.22.91.250 | 63667 | Modbus/T… | 65 |
| 51 | 06:00:07.431947 | 10.22.91.250 | 63667 | 10.22.91.22 | 502 | TCP | 54 |
| 52 | 06:00:07.510294 | 10.22.91.250 | 63667 | 10.22.91.22 | 502 | Modbus/T… | 66 |
| 53 | 06:00:07.511217 | 10.22.91.22 | 502 | 10.22.91.250 | 63667 | TCP | 60 |
| 54 | 06:00:07.511626 | 10.22.91.22 | 502 | 10.22.91.250 | 63667 | Modbus/T… | 65 |
| 55 | 06:00:07.556855 | 10.22.91.250 | 63667 | 10.22.91.22 | 502 | TCP | 54 |
| 56 | 06:00:07.610307 | BeldenIndiaP… | | Spanning-tree… | | STP | 60 |
| 57 | 06:00:07.641586 | ReasonTecnol… | | IEEEI&MSocie_… | | PTPv2 | 60 |
| 58 | 06:00:07.641586 | ReasonTecnol… | | IEEEI&MSocie_… | | PTPv2 | 60 |

Fig. 3. A snapshot of a real-world communication pcap file

Either in real-world or in lab-based experimental set-up, the communication packets captured through Wireshark tool [15] reveal information about the communication flow either in IT or OT endeavour or both. The UTC-formatted source and destination addresses, along with their corresponding ports and protocols, are captured along with the respective packet lengths. Typical protocols are GOOSE, TCP, UDP, Modbus, STP, and many others at the intersection of IT and OT intertwined regimes. Over sixty thousand port numbers are available, and some ports are used specifically for the purpose.

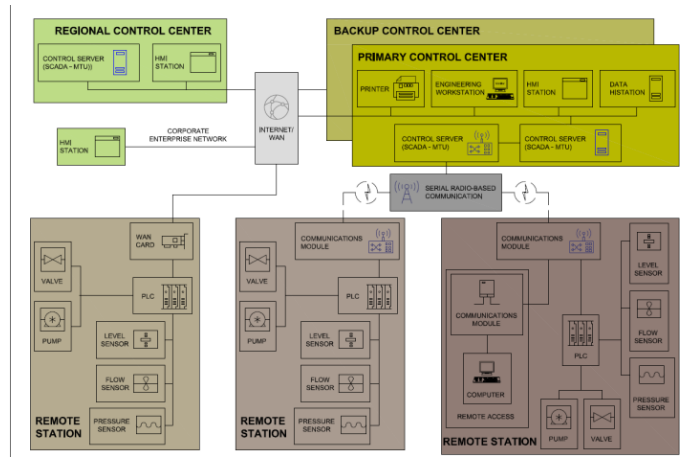

Fig. 4. A typical SCADA system in OT environment [16]

Fig. 4 illustrates the typical SCADA system in OT environment which comprises of many components with remote station and access viz. HMI, sensors, corporate enterprises, etc, are interconnected in a complex way depending upon the requirement. To keep the company network and the OT environment apart, create a DMZ. All communications between the operations management level and the enterprise level must pass through DMZ services. Due to its connections to external environments, the DMZ's services require close monitoring and protection to prevent breaches that would enable attackers to move undetected into the OT environment. But in an OT setting, the digital and the physical are entwined, and there may be a lot of overlap [16].

*C. Cybersecurity aspect of Quantum Communication and SCADA Systems/ Power Networks*

The signcryption scheme, by Ghosh et al. for SCADA [13] using the Quantum Information Toolkit in Python algorithm proposed that the difficulty of factorizing huge prime numbers is the foundation of RSA's security. However, a quantum algorithm can readily solve the factorization problem by applying the superposition principle, in contrast to classical methods. An extensive search using Grover's method in a quantum context takes $O(\sqrt{N})$, whereas a classical version uses $O(N)$. Grover's quantum search technique for prime number factorization weakens symmetric algorithms like AES. [17], [18].

When a quantum system is attempted to be cloned, flaws are introduced into the states of the duplicates. This results from the no-cloning theorem, the foundation of security for quantum communications and a fundamental law of quantum physics. A quantum state can be maximally accurately replicated via a variety of optimal cloning strategies, notwithstanding the prohibition against perfect duplicates. For low-dimensional photonic states, optimal quantum cloning has been experimentally achieved at the edge of the physical limit imposed by the Heisenberg uncertainty principle and the no-signaling theorem. Nonetheless, quantum communication protocols and quantum computation considerably benefit from an increase in the dimensionality of quantum systems [19].

BB84 [5], [20]The quantum key distribution (QKD) scheme, introduced by Bennett and Brassard in 1984 at the Bangalore IEEE conference and later named after them, is secure, as demonstrated separately by Google and Microsoft, with no photon loss and attenuation on fiber. Moreover, accurate laser can emit single photons. QKD through BB84 guarantees key confidentiality but does not resolve the availability problem, as Eve can potentially compromise the fiber.

The photon number splitting (PNS) attack is used in quantum hacking to exploit implementation flaws rather than attack the QKD concepts. Another type of attack is known as a blinding attack, which involves quantum eavesdropping without interception. By delivering a few carefully timed, faint light pulses into the detector's receivers, they can gradually blind them and gain control, whichever is active. Using a Trojan horse attack technique, Eve can transmit signals that can reach Alice's and Bob's properties via quantum channels.

The quantum-key distribution method QKD systems are generally thought to be secure and are capable of transmitting quantum signals over more than 100 kilometers of standard optical fibre [21]. This also led to the research on quantum hacking. Quantum hacking, an experimental example of a successful attack against a commercial quantum key distribution (QKD) system, was demonstrated by Zhao et al. Breaking a well-designed commercial QKD system using solely modern technologies is highly unexpected. The work challenged them to reevaluate the security of workable QKD systems and their real-world applications by demonstrating the slick nature of these systems. They employ the time-shift assault that they suggested earlier. Since Eve does not need to prepare any measurements or states, the time-shift attack is easy to execute [21].

Cybersecurity in power networks falls at the interface or intersection of many conjoined domains [22]. From communication to electrical power systems operations for protecting the grid, it is imperative that any identified application areas for the quantum regime ideally require a broad overview of domain knowledge. However, the core objective is ensuring the availability of power on a 24x7 basis, upholding the other appropriate power system domain metrics.

As cyber attackers' growing sophistication is apparent, it is likely surprising that future cybersecurity breaches would not entail quantum attacks correspondingly. It is expected that the industry cum utility-specific use will eventually have also to take cognizance of quantum computing correspondingly.

As evident from the NIST guidelines, in the quantum computing endeavor, traditional security viz. RSA, ECC, and DSA are no longer secure[23], Table I. While works on post-quantum cryptography are underway, associated research appears need of the hour.

TABLE I. IMPACT OF QUANTUM COMPUTING ON COMMON CRYPTOGRAPHIC ALGORITHMS [23]

| Cryptographic Algorithm | Type | Purpose | Impact from large-scale quantum computer |
|---|---|---|---|
| AES | Symmetric Key | Encryption | Larger key sizes are needed |
| SHA-2, SHA-3 | -------- | Hash functions | Larger output needed |
| RSA | Public Key | Signatures, Key establishment | No longer secure |
| ECDSA, ECDH | Public Key | Signatures, Key exchange | No longer secure |
| DSA | Public Key | Signatures, Key exchange | No longer secure |

The first quantum algorithms that applied to issues outside of the realm of quantum mechanics achieved a super-polynomial speedup over classical algorithms and had clear applications were Shor's algorithms [24]. Specifically, Shor's algorithms could be utilized to the cryptosystem of RSA [25]. Given the difficulty of factoring integers that are the product of two primes of comparable sizes (henceforth referred to as "RSA integers"), Gidney et al. (2021) [26] factored n-bit RSA integers of 2048 bits using $3n+0.002n \lg n$ logical qubits, $0.3n^3 + 0.0005n^3 \lg n$ To_olis, and $500n^2 + n^2 \lg n$ measurement depth. RSA is broken using the Shor algorithm in 8 hours with 20 million noisy qubits.

This paper examines the plausibility of scenarios to identify the scope, opportunities, and challenges of quantum communication and computing endeavors, and proposes a model for distributed SCADA networks that utilizes quantum walks, QKD, error correction, and eavesdropping prevention. The contribution of this paper is to use an open-source testbed experimental SCADA cybersecurity dataset using a Python environment based on Google Colab [27] with IBM qiskit [28], [29] SDK based quantum circuit development and implementation. The novelty lies in proposing an integrated quantum approach that builds circuits for distributed SCADA quantum walks, QKD, and combines both, incorporating features from an open-source dataset of SCADA cybersecurity. This approach addresses generalization issues for potential broader implementation.

### III. QUANTUM APPROACH OF SCADA COMMUNICATION

The approach in this work entails selecting a SCADA cybersecurity processing dataset, developing a quantum walk-based circuit for various SCADA centers, generating QKD circuits, and finally integrating these components to emulate a real-world simulation. The details are provided in the subsections below.

*A. SCADA Dataset selection*

WUSTL open source dataset [30] details are available in the citation link. Table II depicts the variables and sample 5 rows of the data. A sample row has been considered for the demonstration of this work. The source port, total packets, total bytes, source packets, destination packets and source bytes variables considered for the row-wise quantum data encoding. Though the dataset is labeled data, however, the class variable have not been used as the purpose is not building quantum classification however, methods of quantum communication through the circuits.

TABLE II. DATASET SAMPLE ROWS AND FEATURES

|  | Sport | TotPkts | TotBytes | SrcPkts | DstPkts | SrcBytes |
|---|---|---|---|---|---|---|
| **100** | 55036 | 18 | 1152 | 10 | 8 | 644 |
| **101** | 55037 | 18 | 1152 | 10 | 8 | 644 |
| **102** | 55041 | 20 | 1276 | 10 | 10 | 644 |
| **103** | 55039 | 20 | 1276 | 10 | 10 | 644 |
| **104** | 55040 | 20 | 1276 | 10 | 10 | 644 |

*B. Quantum Walk for Distributed SCADA*

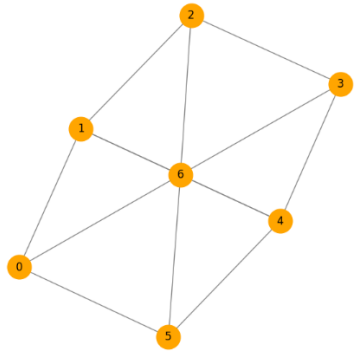

Fig. 5. SCADA networks with the central node connecting all other nodes

IBM computing endeavour [31] offers composer/ simulator and quantum processor for qiskit implementation of circuits. Using this, quantum circuits have been developed. Fig. 5 depicts a topology of the six SCADA centers, all connected to the central node, which is considered the RTU/load dispatch center. The others are also associated with the neighbors. The topology is quantum walk based depicted by the graph where the nodes represent the SCADA centers and the edge as communication links. This aims to emulate a small-scale real-world scenario where transmission networks are connected from various parts for power evacuation. Further, quantum feature encoding is followed based on the SCADA dataset. The circuit, Fig. 6 corresponding to Fig. 5, comprises of a deep architecture of Hadamard gates, phase flips and multi-cubit entanglements.

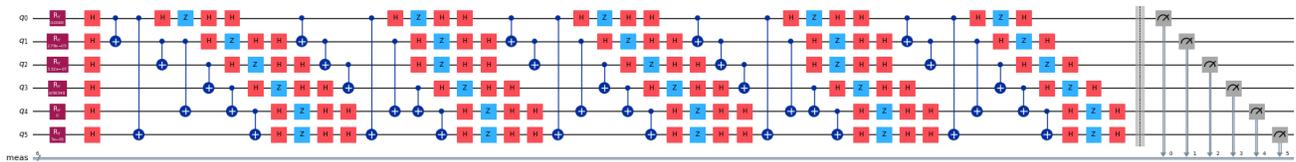

Fig. 6. Quantum circuit of the SCADA networks

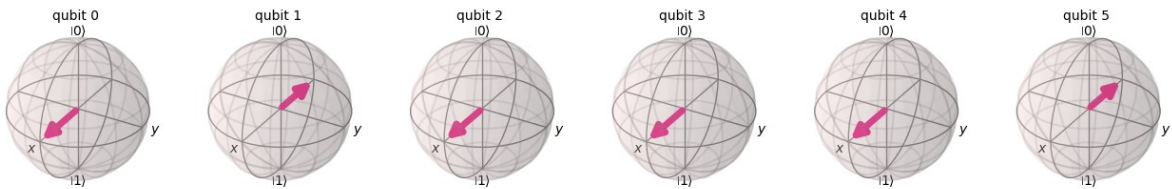

Fig. 7. Multivector plot of quantum circuit

Fig. 7 depicts the multivector bloch sphere where the pink arrow indicates the corresponding states. The indication reveals more or less the same direction i.e., correlated in-between state |0⟩ and |1⟩ tilting in a superposition of states.

A closer view of Fig. 5 depicts the SCADA dataset sample row of six variables encoded in the six qubits, Fig. 8. The quantum data encoding techniques in qubits [32] is done by normalizing data followed by the Hadamard gate and C-NOT operations.

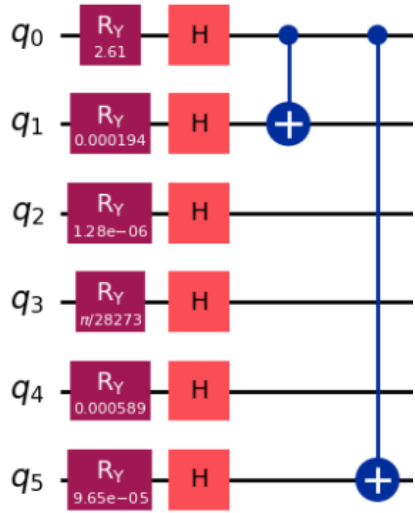

Fig. 8. Data encoded Qubits sample row of SCADA dataset

-0.1254214318 | 000000)-0.1254302133 | 000001)+0.12542496 | 000010)+0.1254337418 | 000011)-0.1245718445 | 000100)-0.1245805665 | 000101)+0.1245752613 | 000110)+0.1245839835 | 000111)-0.1254179477 | 001000)-0.125426729 | 001001)+0.1254145078 | 001010)+0.1254232889 | 001011)-0.1245683841 | 001100)-0.1245771059 | 001101)+0.1245648799 | 001110)+0.1245736015 | 001111)-0.12542496 | 010000)-0.1254337418 | 010001)+0.1254214318 | 010010)+0.1254302133 | 010011)-0.1245752613 | 010100)-0.1245839835 | 010101)+0.1245718445 | 010110)+0.1245805665 | 010111)-0.1254145078 | 011000)-0.1254232889 | 011001)+0.1254179477 | 011010)+0.125426729 | 011011)-0.1245648799 | 011100)-0.1245736015 | 011101)+0.1245683841 | 011110)+0.1245771059 | 011111)+0.1254144638 | 100000)+0.1254232448 | 100001)-0.1254179918 | 100010)-0.1254267731 | 100011)+0.1245649237 | 100100)+0.1245736452 | 100101)-0.1245683403 | 100110)-0.1245770621 | 100111)+0.1254249159 | 101000)+0.1254336977 | 101001)-0.1254214758 | 101010)-0.1254302574 | 101011)+0.124575305 | 101100)+0.1245840273 | 101101)-0.1245718007 | 101110)-0.1245805227 | 101111)+0.1254179918 | 110000)+0.1254267731 | 110001)-0.1254144638 | 110010)-0.1254232448 | 110011)+0.1245683403 | 110100)+0.1245770621 | 110101)-0.1245649237 | 110110)-0.1245736452 | 110111)+0.1254214758 | 111000)+0.1254302574 | 111001)-0.1254249159 | 111010)-0.1254336977 | 111011)+0.1245718007 | 111100)+0.1245805227 | 111101)-0.124575305 | 111110)-0.1245840273 | 111111)

The quantum state represents the superposition of the 6-qubit system in the circuit, where each coefficient corresponds to an amplitude. The sum of the squares of these amplitudes represents the probability, approximately equal to 1. Fig. 9 depicts the probability distribution bar chart of these amplitudes.

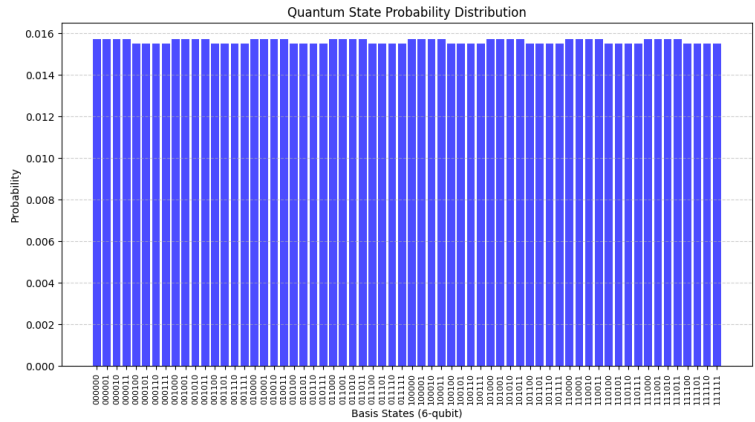

Fig. 9. Probability histogram of the quantum circuit states

## C. QKD for SCADA

Separately, Fig. 10 illustrates the QKD circuit, comprising six qubits corresponding to the six nodes of the SCADA walk, with the application of Hadamard gates for superposition, X-gates for bit-flip operations, and unitary gates for phase rotation. The measurement gates collapse the state into the classical state for observation.

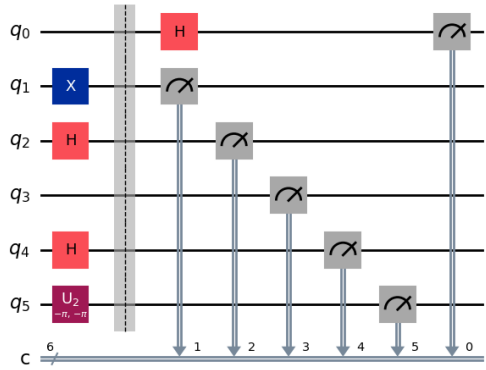

Fig. 10. A quantum circuit for QKD

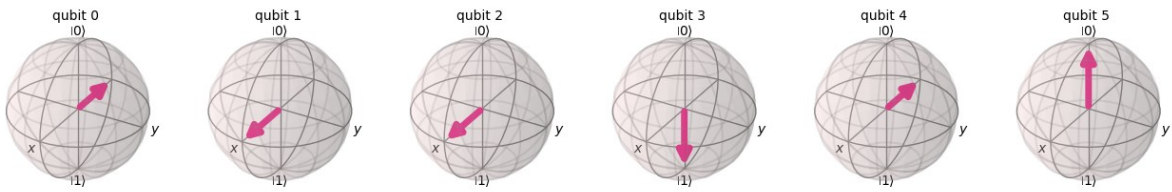

Fig. 11. Multivector plot of QKD circuit

Fig. 11 depicts the QKD state of six qubits, caused by the gate operations and apparently diverse i.e., entanglement and rotation operations which is in contrast with Fig. 7 multivector plots. However, neither of these indicate any GHZ state since these not appear to be in mixed state.

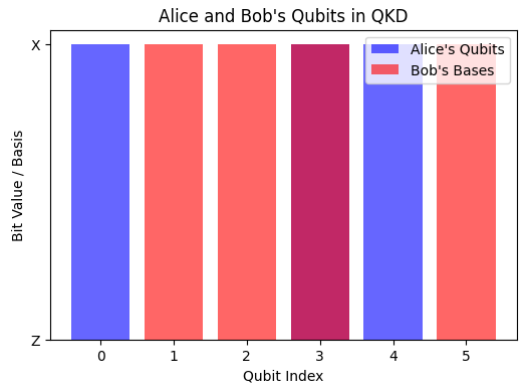

Fig. 12. Qubit index and basis plot of all six qubits in qiskit [6]

Fig. 12 depicts the BB84 QKD process, where Alice sends the quantum bits, and Bob randomly bases measure them correspondingly from qubits indexed 0 to 5. The bit value in the ordinate indicates the choice of basis, i.e., either X or Z. If both Alice and Bob choose the Z or X basis correspondingly, this leads to correct measures; otherwise, it results in incorrect measures. The presence of Eve, the eavesdropper, will introduce errors leading to the collapse of the state. Fig. 13 illustrates the quasi-probability distribution of prime numbers along the abscissa, resulting from QKD random generation corresponding to specific bit positions or indices.

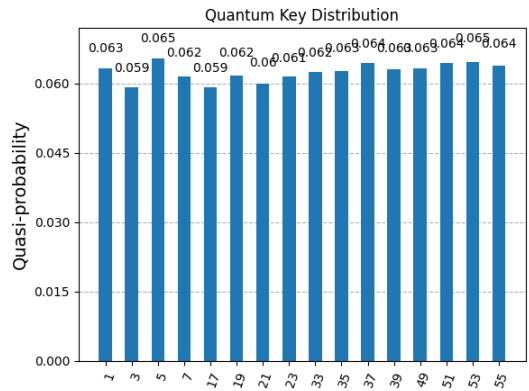

Fig. 13. QKD quasi-probability reproduced in qiskit

*D. Integrated Approach for Quantum Walk and QKD*

Fig. 14 illustrates the error rate in QKD over multiple sessions, which are monitored for possible eavesdropping. Each session corresponds to a specific QKD exchange between Alice and Bob. The threshold value above indicates potential eavesdropping and the red line indicates the corresponding change in error rate over the session.

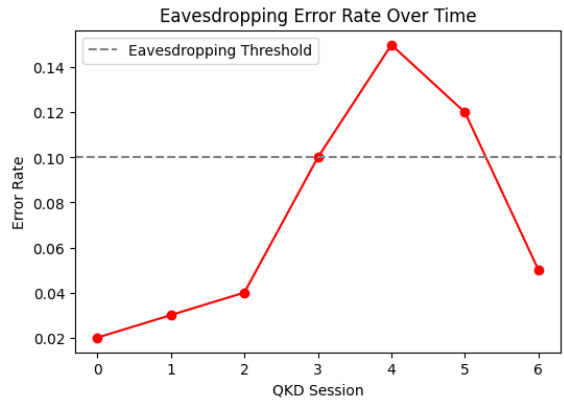

Fig. 14. Eavesdropping rate on QKD session over a thresold values reproduced in qiskit

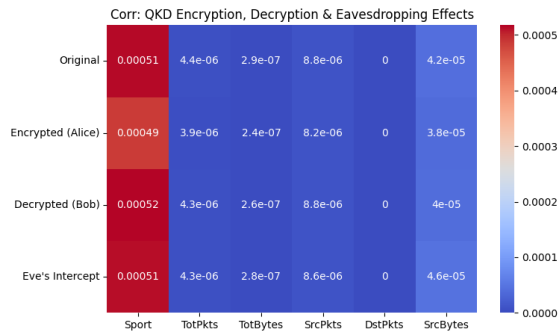

Fig. 15. Heatmap showing the correlation among QKDs

The heatmap in Fig. 15 reveals the correlation coefficients over the variables i.e., Source port, Total packets, Total bytes, source packets, destination packets and source bytes with the corresponding original encrypted, decrypted and eavesdropping attempt. The source ports apparently depict higher vulnerability where Eve activity appears intense which by quantum bit error rate (QBER) through may be detectable by Alice and Bob.

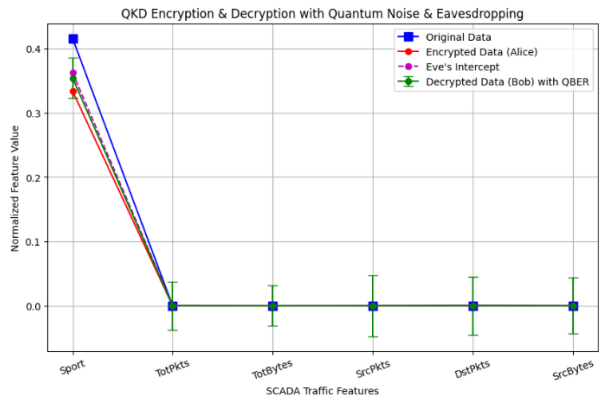

Fig. 16. OKD Encryption and Decryption with Quantum Noise and Eavesdropping

Fig. 16 illustrates the QKD encryption and decryption, along with the corresponding eavesdropping plot, over the six variables of the SCADA dataset, as shown on the abscissa. The variation in the source port indicates a potential eavesdropping attempt while the different color-coded lines are converging, indicating the system is under control. This reveals the possible extent of maintaining integrity during quantum communication. A successful key matching, Fig. 17, indicates secure SCADA communication plausibility where high success rate is achievable by adopting appropriate quantum circuits and gates through different conceptual architecture comprising of many aspects.

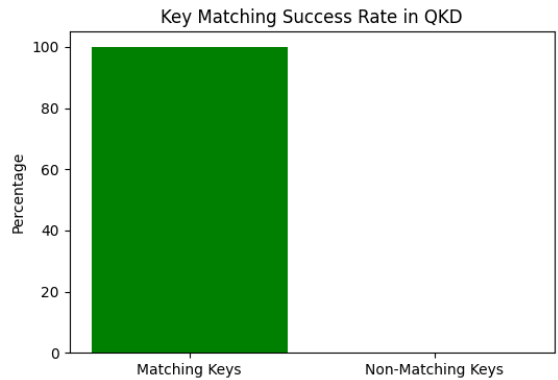

Fig. 17. Quantum key matching success rate comparison

Table III displays a sample data row from the SCADA dataset's original traffic flow, comprising six variables and their corresponding encryption and decryption data. The same Alice and Bob key after the error correction of six qubits reveals the successful quantum operation. The six digit key encourages in real-world scenario implementation of larger bits of 128 or 256 bits matching with the use case.

TABLE III. Sample Original, Encrypted, Decrypted Data AND Keys

| iLoc[6] | |
|---|---|
| **Original Traffic Data:** | [0.4157021167143268, 8.400277651282369e-05, 5.669178141799643e-07, 4.421179216920737e-05, 0.00023442261709409723, 3.986722974382581e-05] |
| **Encrypted Traffic Data:** | [0.5842978832856732, 0.9999159972234872, 0.9999994330821859, 0.9999557882078308, 0.999765577382906, 0.9999601327702562] |
| **Decrypted Traffic Data:** | [0.4157021167143268, 8.400277651277044e-05, 5.669178141376463e-07, 4.421179216917981e-05, 0.00023442261709405443, 3.98672297438063e-05] |
| **Alice's Key:** | [111111] |

| Bob's Key after Error Correction: | [111111] |

These typical data analytics have been performed using Qiskit, which demonstrates that if 'democratized' quantum computing resources become available, operating and controlling electrical grid networks might, at some point, transition from classical and conventional computing paradigms. In this context, it is also noteworthy to mention that quantum communication might need to address the vast complexity of the sophisticated domain day-to-day regime in whatever way, i.e., either through quantum-classical or quantum-quantum paradigms deemed fit.

## IV. Discussion

The complexity of current communication in power networks necessitates the installation of optical fiber ground wire, typically ranging from 24-core to 96-core, over long-distance transmission lines, whether they are overhead or underground cables. The OPGW terminates in the substation wherein, through a cable trench, it is laid within the substation yard and control room. In the control room, Ethernet connections are made through the substation automation system communication panel and are connected to other panels, such as the transformer, bus coupler, and others, depending on the substation's size and complexity. The complexity further increases when adhering to the respective monitoring, control, protection, and communication industry standards, which are typically followed by ITU's various series.

Packet capture and subsequent data analytics reveal the source address, destination address, source ports, and protocols used in the system's communication regime.

Utilizing the IEC 61850 standard for a 220/33 kV substation. The accompanying logical nodes, logical devices, and corresponding functions—control, protection, or measurement—for power, protection, and monitoring reasons are also revealed by further examination of Goose PDU. Seven messages match the message type and timing as specified in IEC 61850. A comprehensive data analytics approach can be used to visualize the mapping with the signal list, including the bays, bay control unit (BCU) for the relevant voltages, circuit breaker activities, and isolators for opening/closing. Given that the substation has multiple bays, as indicated in the schematic diagram, and numerous IEDs based on voltage class, data visualization and accompanying lengths can provide helpful information for further analysis. Despite the speed at which Wireshark records streaming data, which makes it challenging, if not impossible, to follow without the use of filtering techniques manually, data analytics can offer insight into the substation operation at various points.

As evident from the related works, quantum communication through optical fiber has been researched in terms of capacity, Efficiency, and quantum-structured light. However, when addressing the complexity of real-world power network scenarios, SCADA communication as explained above, should also be considered for plausible domain intersections.

The above works demonstrate that quantum circuits are designed for distributed SCADA networks comprising 6 centers, emulating a central one that resembles a remote terminal unit (RTU) or load dispatch centre for a typical power network, and are encoded with the features of an open-source cybersecurity operations technology (OT) dataset. Moreover, another quantum circuit for QKD, based on the BB84 protocol, has been developed and later integrated with the first circuit. The encrypted data using the quantum key exchange and decrypted data sample were compared, and the corresponding keys of Alice and Bob were tabulated. Therefore, the following opportunities and challenges are envisaged.

**Opportunity**

i) Eavesdropping can be detected and shines fresh thought and inspiration in traditional cybersecurity endeavors.

ii) Traditional encryption algorithms will likely face an existential crisis in the quantum computing landscape, but QKD offers a plausible and sustained opportunity.

iii) Quantum computing coupled with quantum communication can offer inspiration for potentially unreachable computing areas of classical computers, viz. sustainability, cybersecurity, and related domains.

iv) Quantum machine learning can address the possible saturation of hardware and software trade-offs in terms of performance computing resources and time.

v) As power system communication is predominantly optical fiber-based through overhead transmission lines and substations at different voltages, optical fiber-based quantum communication would likely fit in.

vi) Quantum number generation can also be a right fit in other related areas of SCADA communication.

**Challenges**

i) SCADA in power networks is complex, depending on their size. The standards used are manifold for power system operation and corresponding cybersecurity in the IT-OT interwoven regime. The most relevant series are IEC 61850, IEC 60870, IEC 62351, IEC 62443, and ITU G.652 series and many others.

ii) The essence of a smart grid is the bi-directional simultaneous flow of power and communication without compromising with other domains.

iii) The usual IT domain confidentiality, Integrity, and availability (CIA triad) stands in the reverse order, i.e., AIC, making availability the key priority for meeting customer requirements.

iv) In the Power system SCADA operations OT endeavors, protocols are specific viz. GOOSE [31], MMS, IEC 101/104, and many others where usually it is contemplated that data training is not required as it is very specific communication purposed for IED-IED and corresponding RTU/Load despatch center

v) Network packet sniffing is a procedure that is usually maintained, among other things, viz. Wireshark [15], a tool usually deployed

vi) The intrusion detection and protection system (IDPS) typically employs both signature- and anomaly-based methods. However, most SIEM tools utilize a combination of these.

vii) The vast amount of real-time data generated by power system operations needs real-time situational awareness.

In real-world SCADA scenarios that adhere to appropriate standards, numerous protocols, IP addresses, and MAC addresses of devices for power system monitoring, control, and operations stream live data at an overwhelming rate. The typical single sample of the dataset exploration and results provides an encouraging indication for further exploration in the corresponding domain intersection.

## V. Conclusion

As seen from the above, the paper attempts to consolidate the plethora of research works in quantum communication and computing that are relevant to the communication and cybersecurity paradigm of power networks, one of a nation's critical infrastructures. It cites research works in the quantum communication domain, focusing on existing communication systems, standards, and protocols for power networks and their relationship to cybersecurity paradigms. As optical fiber-based communication for power networks is a time-tested tool, the already mature technology, and practice should also be gradually prepared for broad acceptance in the quantum era by utility personnel who operate, manage, and protect the grids. Therefore, disseminating the current research flavor should also be a practical reality. Quantum computing have been performed to demonstrate the plausible future of the working domain in the coming days as adjoining domain areas viz. quantum neural networks for malicious data classification and data encoding complexity are some of the issues need addressing in real-world scenario [33], [32]. As the sophistication in the classical regime unfolds, so does the corresponding research dissemination in the quantum realm across the communities. Should a quantum computer transpire, along with the corresponding communication and computing paradigm, how prepared is the real world of power networks communication for accepting the plausible reality and coping with the new regime from different domains? An intersection area of interest and focus has been addressed in this paper. Applying quantum machine learning techniques to the existing classical dataset for a SCADA large distributed networks in this overlapping domain has been identified as a future direction.

## Acknowledgment

I acknowledge the support and encouragement of everyone who enabled us to develop this article.